\documentclass[5p,sort&compress,times,letterpaper]{elsarticle}
\usepackage{amsmath,amssymb}
\usepackage{bm}
\usepackage{graphicx}
\usepackage{color}
\usepackage[utf8]{inputenc}

\newcommand{\rmi}{\mathrm{i}}
\newcommand{\rmd}{\mathrm{d}}
\newcommand{\km}{\mathrm{km}}
\newcommand{\s}{\mathrm{s}}
\newcommand{\MeV}{\mathrm{MeV}}
\newcommand{\erg}{\mathrm{erg}}
\newcommand{\mus}{\mu\mathrm{s}}
\newcommand{\tot}{\mathrm{tot}}
\newcommand{\tr}{\mathrm{tr}}
\newcommand{\pfrac}[2]{\left(\frac{#1}{#2}\right)}

\newcommand{\Gf}{G_\mathrm{F}}

\newcommand{\sfH}{\mathsf{H}}
\newcommand{\Hvac}{\mathsf{H}_\text{vac}}

\newcommand{\Hmat}{\mathsf{H}_\text{mat}}
\newcommand{\Hnu}{\mathsf{H}_{\nu\nu}}
\newcommand{\Pe}{\varrho_{ee}}
\newcommand{\Pt}{\varrho_{xx}}
\newcommand{\bhv}{\boldsymbol{\hat{\mathbf{v}}}}
\newcommand{\bnabla}{\boldsymbol{\nabla}}
\newcommand{\bfx}{\mathbf{x}}
\newcommand{\bfp}{\mathbf{p}}

\newcommand{\kmax}{\kappa^\text{max}_\varpi}

\begin{document}

\title{Neutrino flavor instabilities in a time-dependent supernova model}

\author{Sajad Abbar}
\author{Huaiyu Duan
}
\address{Department of Physics \& Astronomy, 
University of New Mexico,
Albuquerque, NM 87131, USA}
\date{\today}

\begin{abstract}

A dense neutrino medium such as that inside a core-collapse supernova can
experience collective flavor conversion or oscillations because of the
neutral-current weak interaction among the neutrinos. This phenomenon has
been studied in a restricted, stationary supernova model which possesses
the (spatial) spherical symmetry about the center of the supernova and the
(directional) axial symmetry around the radial direction. Recently it
has been shown that these spatial and directional symmetries
can be broken spontaneously by collective neutrino oscillations. In
this letter we 
analyze the neutrino flavor instabilities in a time-dependent supernova
model. Our results show that collective neutrino oscillations
start at approximately the same radius in both the stationary and
time-dependent supernova models unless there exist very rapid
variations in local physical conditions on timescales of a few
microseconds or shorter. Our 
results also suggest that collective neutrino oscillations can vary
rapidly with time in the regimes where they do occur which need to
be studied in time-dependent supernova models.

\end{abstract}
 
\maketitle

\section{Introduction}
 
Neutrinos are essential to the thermal, chemical and dynamical
evolution of the early universe and some of the compact objects such
as the proto-neutron star inside a core-collapse supernova (SN). Whenever
there is a difference between the energy spectra and/or fluxes of the
electron-flavor neutrino/antineutrino and other neutrino species, the flavor
conversion or oscillations among different neutrino flavors can also
have important impacts on  nucleosynthesis and other physics
inside these hot and dense astrophysical environments.

In the absence of collision the flavor evolution of the neutrino obeys
the Liouville equation
\cite{Sigl:1992fn,Strack:2005ux,Cardall:2007zw} 
\begin{align}
\label{eq:eom-full}
\partial_t\rho + \bhv\cdot\bnabla\rho
&= -\rmi[\Hvac + \Hmat + \Hnu,\, \rho],
\end{align}
where $\bhv$ is the velocity of the neutrino, $\rho(t,\bfx,\bfp)$ is
the (Wigner-transformed) flavor density matrices of the neutrino which
depends on time 
$t$, position $\bfx$ and neutrino momentum $\bfp$, $\Hvac$ is the
standard vacuum Hamiltonian,  and $\Hmat$ and $\Hnu$ are
the matter and neutrino potentials, respectively. 
The neutrino potential in Eq.~\eqref{eq:eom-full} takes the following form
\cite{Fuller:1987aa,Notzold:1987ik,Pantaleone:1992eq}: 
\begin{align}
\sfH_{\nu\nu} = \sqrt2\Gf\int\frac{\rmd^3 p'}{(2\pi)^3}(1-\bhv\cdot\bhv')
[\rho(t,\bfx,\bfp')-\bar\rho(t,\bfx,\bfp')],
\label{eq:Hvv}
\end{align}
where $\Gf$ is the Fermi coupling constant, and $\bar\rho$ is the
density matrix of the antineutrino. 
Because the neutrino potential couples neutrinos of
different momenta,  a dense neutrino medium can oscillate in a
collective manner (see, e.g.,
\cite{Kostelecky:1993yt,Pastor:2001iu,Pastor:2002we,Abazajian:2002qx,Balantekin:2004ug,Duan:2006jv,Duan:2006an,Duan:2007sh,Raffelt:2007cb,Fogli:2007bk,Dasgupta:2008my,Dasgupta:2008cu,Dasgupta:2009mg,Gava:2009pj,Friedland:2010sc,Duan:2010bf,Pehlivan:2011hp,Cherry:2012zw,deGouvea:2012hg,Malkus:2014iqa};
see also \cite{Duan:2010bg} for a review). 

Eq.~\eqref{eq:eom-full} poses a challenging 7-dimensional problem (not taking
into account the dimensions in neutrino flavors), and it has never been
solved in its full form. In previous studies various 
simplifications  have been made so that a numerical or analytic
solution to this equation can be found. For neutrino oscillations in
SNe a commonly used model is the (neutrino) 
Bulb model \cite{Duan:2006an}. In this model a spatial spherical
symmetry around the center of the SN is imposed so that it has only one
spatial dimension. An additional directional axial symmetry
around the radial direction is imposed to make the model
self-consistent which reduces the number of momentum dimensions to two.
One also imposes the time translation symmetry because the timescale
of neutrino oscillations is much shorter than those in neutrino
emission or dynamic evolution in SNe.
However, it has been shown in a series of recent studies that both the
spatial and directional symmetries can be 
broken spontaneously by collective neutrino oscillations if they are
not imposed 
\cite{Raffelt:2013rqa,Mirizzi:2013rla,Duan:2013kba,Mangano:2014zda,Duan:2014gfa,Mirizzi:2015fva,Mirizzi:2015hwa,Chakraborty:2015tfa,Abbar:2015mca} 
(see also \cite{Duan:2015cqa} for a short review). 
In both cases small deviations from the initial symmetric conditions
are amplified by the symmetry-breaking oscillation modes which 
can occur closer to the neutrino
sphere than the symmetry-preserving modes do.

It is natural to wonder if collective neutrino oscillations can also break
the time-translation symmetry spontaneously in SNe
\cite{Raffelt:2013rqa}. If they do, 
then a time-independent SN model may not accurately describe the
neutrino oscillation phenomenon in SNe even though the typical
timescale of the variation in the 
neutrino emission is much longer than that of neutrino
oscillations. In this letter we analyze the neutrino flavor stability
in a time-dependent SN model which should provide some interesting insights to
this question.

\section{Time-dependent neutrino Bulb model}

\begin{figure} 
\begin{center}
\includegraphics*[width=.45\textwidth]{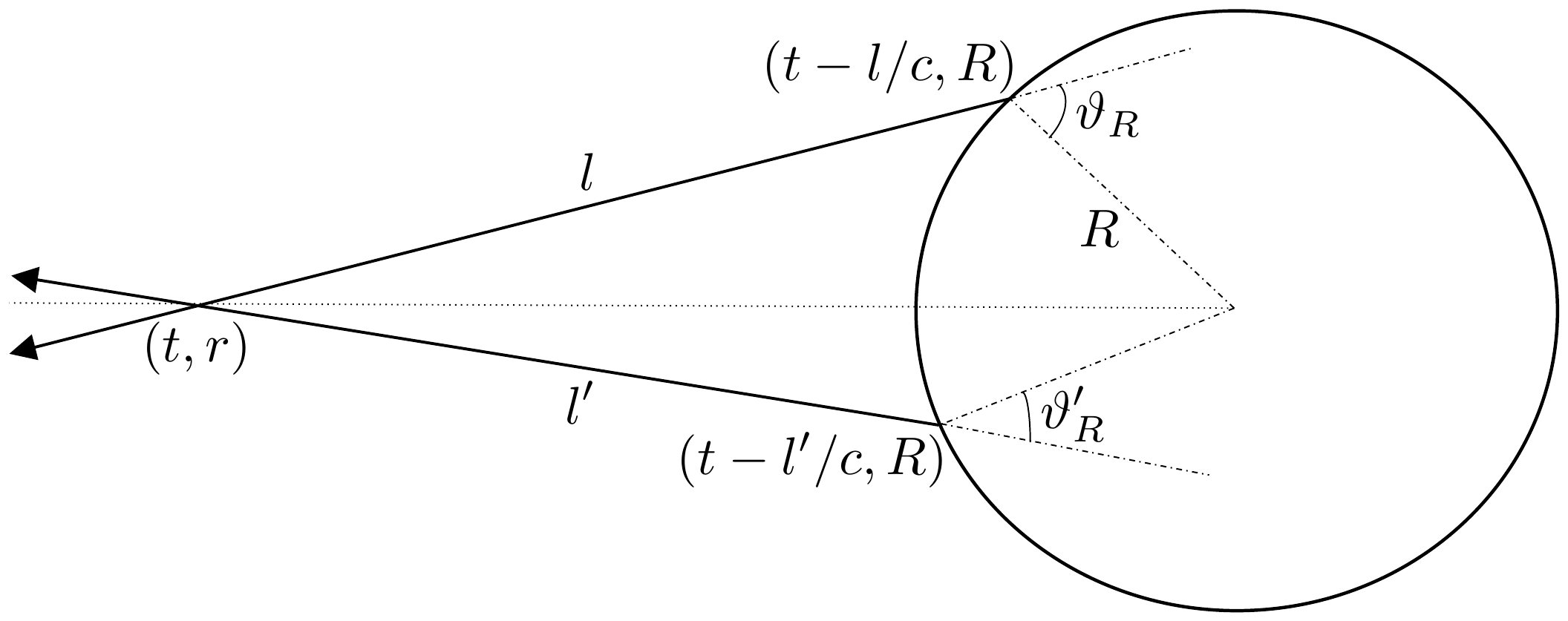}
\end{center}
\caption{The geometric picture of the time-dependent (neutrino) Bulb model for
  supernova. Two neutrinos emitted from the neutrino sphere of radius
  $R$ with emission angles $\vartheta_R$ and $\vartheta'_R$ and at time
  $t-l/c$ and $t-l'/c$ meet each other at radius $r$ and time $t$,
  where $l$ and $l'$ are the distances by which the two neutrinos have
  traveled
  from the neutrino sphere to their meeting point, respectively.
}
\label{fig:bulb}
\end{figure}

We will focus on the potential differences between the results obtained
from the time-dependent and stationary SN models. Therefore, we will
employ the time-dependent Bulb model which has the same spatial
spherical symmetry and the directional axial 
symmetry as in the conventional Bulb model. 
Unlike the conventional stationary Bulb model, however,
we will not assume that the emission and flavor evolution of the
neutrinos are time-independent (see Fig.~\ref{fig:bulb}). For
simplicity, we will  
consider the mixing between two active flavors, the $e$
and $x$ flavors, with the latter being the linear superposition of the
$\mu$ and $\tau$ flavors. We also assume a small vacuum mixing angle
$\theta_\mathrm{v}\ll1$. 

It is convenient to use the vacuum oscillation frequency
\begin{align}
  \omega = \pm \frac{|\Delta m^2|}{2 E}
\end{align}
to label the neutrino and antineutrino with energy $E$, where
$\Delta m^2$ is the neutrino mass-squared difference, and the
plus and minus signs apply to the neutrino and the antineutrino,
respectively. We define reduced neutrino density matrix
\begin{align}
  \varrho(t;r;\omega,u) \propto
  \left\{\begin{array}{ll}
  \rho & \text{ if } \omega > 0, \\
  \bar\rho & \text { if } \omega < 0
  \end{array}\right.
\end{align}
with normalization 
\begin{align}
  \tr\varrho = 1,
\end{align}
where $u=\sin^2\vartheta_R$ with
$\vartheta_R$ being the emission angle of the neutrino on the
neutrino sphere (see Fig.~\ref{fig:bulb}), and $r$ is the radial
distance from the center of the SN.

The equation of motion (EoM) for the (reduced) density matrix
$\varrho$ can be written as
\begin{align}
  \rmi(\partial_t+ v_u\partial_r)\varrho
  =[\Hvac + \Hmat + \Hnu,\, \varrho],
  \label{eq:eom}
\end{align}
where 
\begin{align}
v_u(r) = \sqrt{1-\pfrac{R}{r}^2 u}
\end{align}
is the radial velocity of the neutrino. In the weak interaction basis
the standard vacuum Hamiltonian and the
matter potential are
\begin{align}
  \Hvac &\approx  -\frac{\eta\omega}{2} \sigma_3
  =-\frac{\eta\omega}{2} \begin{bmatrix}
    1 & 0 \\ 0 & -1 \end{bmatrix}  \\
\intertext{and}
  \Hmat & = \lambda \begin{bmatrix} 
    1 & 0 \\ 0 & 0 
  \end{bmatrix}
                 = \begin{bmatrix}
                   \sqrt2\Gf n_e & 0 \\ 0 & 0 
\end{bmatrix},
\end{align}
respectively,
where $\eta=+1$ and $-1$ for the normal (neutrino mass) hierarchy (NH,
i.e.\ with $\Delta m^2 > 0$) and the inverted hierarchy (IH,
$\Delta m^2 < 0$), respectively, and $n_e$ is the net electron number density.

In this letter we assume that 
the number flux $F_{\nu_\alpha/\bar\nu_\alpha}(E,\vartheta_R)$ of the
neutrino/antineutrino in flavor $\alpha$ ($\alpha=e,x$)
is time independent \cite{Mirizzi:2011tu}. We define
the distribution function of the neutrino emission to be
\begin{align}
  g(\omega,u) \propto \left|\frac{\rmd E}{\rmd \omega}\right|\times
  \left\{\begin{array}{ll}
  (F_{\nu_e}+F_{\nu_x}) & \text{ if } \omega > 0, \\
  -(F_{\bar\nu_e}+F_{\bar\nu_x}) & \text{ if } \omega < 0
  \end{array}\right.
\label{eq:g}
\end{align}
with normalization conditions 
\begin{subequations}
\begin{align}
  \int_0^\infty\rmd\omega \int_0^1\frac{\rmd u}{2}\, g(\omega,u) &= 1, \\
  \int_{-\infty}^0\rmd\omega \int_0^1\frac{\rmd u}{2}\, g(\omega,u) &=
  -\frac{N_{\bar\nu}^\tot}{N_\nu^\tot},
\end{align}
\end{subequations}
where 
\begin{subequations}
\begin{align}
  N_\nu^\tot &= \int_0^\infty\rmd E\int_0^1\frac{\rmd u}{2}\, 
  (F_{\nu_e}+F_{\nu_x}),\\
  N_{\bar\nu}^\tot &= \int_0^\infty\rmd E\int_0^1\frac{\rmd u}{2}\,
  (F_{\bar\nu_e}+F_{\bar\nu_x})
\end{align}
\end{subequations}
are the total number luminosities of the neutrino and antineutrino
(i.e.\ the number of neutrinos or antineutrinos emitted by the whole
neutrino sphere per unit time), respectively.
The opposite signs of $g(\omega,u)$ for the neutrino and antineutrino
in Eq.~\eqref{eq:g}
take into account their different contributions to the neutrino
potential in Eq.~\eqref{eq:eom-full}.
In the Bulb model the neutrino potential can be written as
\begin{align}
  \Hnu(t;r;u) 
  &= \frac{\sqrt2\Gf N_\nu^\tot}{4\pi r^2} 
  \int_{-\infty}^\infty\rmd\omega'
  \int_0^1\frac{\rmd u'}{v_{u'}}\,(1-v_u v_{u'})
\nonumber\\
  &\quad\times g(\omega',u')\,\varrho(t;r;\omega',u').
\end{align}
Because collective neutrino oscillations usually occur in the regime
$R/r\ll1$ in the Bulb model, we will take the large-radius approximation \cite{EstebanPretel:2008ni}
\begin{align}
v_u(r) \approx 1-\pfrac{R}{r}^2\frac{u}{2}.
\end{align}
In this approximation,
\begin{align}
\Hnu(t;r;u) \approx
\mu \int \pfrac{u+u'}{2} g' \varrho'\,\rmd\Gamma',
\label{eq:Hnu}
\end{align}
where all the primed quantities are functions of $u'$ and $\omega'$, 
e.g., $\varrho'=\varrho(t;\omega',u'; r)$,
\begin{align}
\mu(r) =  \frac{\sqrt2\Gf N_\nu^\tot}{4\pi R^2}\pfrac{R}{r}^4
\label{eq:mu}
\end{align}
is the strength of the neutrino potential at radius $r$, and
\begin{align}
\int\rmd\Gamma' \equiv \int_{-\infty}^\infty\rmd\omega'
\int_0^1\rmd u' .
\end{align}

\section{Linear regime}

In the regime where no significant flavor transformation has occurred, the
linear flavor-stability analysis is applicable
\cite{Banerjee:2011fj}. In this regime the neutrino density 
matrices take the form
\begin{align}
\varrho(t;r;\omega,u)\approx 
\frac{\Pe+\Pt}{2}
\begin{bmatrix}
1 & 0 \\ 0 & 1 
\end{bmatrix}
+\frac{\Pe-\Pt}{2}
\begin{bmatrix}
1 & \epsilon \\ \epsilon^* & -1 
\end{bmatrix},
\end{align}
where $\Pe(\omega,u)$ and $\Pt(\omega,u)$ are the probabilities
for the neutrino (or antineutrino) to be in the $e$ and $x$
flavors, respectively, and $|\epsilon(t;r;\omega,u)|\ll1$.
Here in the spirit of flavor-stability analysis we have assumed that
$\varrho$ are approximately constant. At the onset of collective
neutrino oscillations $\epsilon$ grow exponentially. If
$\epsilon$ has strong time dependence, the time translation
symmetry is broken spontaneously by collective neutrino oscillations.

Keeping only the terms up to $\mathcal{O}(\epsilon)$ in
Eq.~\eqref{eq:eom} we obtain
\begin{align}
\rmi(\partial_t + v_u \partial_r)\,\epsilon
&\approx(-\eta\omega+\lambda +  C)\,\epsilon
\nonumber\\
&\quad-\frac{\mu}{2} \int (u+u')(\Pe'-\Pt')g'\,\epsilon'\,\rmd\Gamma',
\label{eq:eom-lin}
\end{align}
where 
\begin{align}
C(u,\mu) = \frac{\mu}{2} \int (u+u')(\Pe'-\Pt')g' \,\rmd\Gamma'.   
\end{align}
Defining
\begin{align}
\epsilon_\varpi(r;\omega,u) = \int_{-\infty}^\infty
\epsilon(t;r;\omega,u)\,e^{\rmi\varpi t}\,\rmd t
\end{align}
we can rewrite Eq.~\eqref{eq:eom-lin} in frequency space as
\begin{align}
\rmi  \partial_r\epsilon_\varpi
&\approx
  \left\{-\eta\omega+(\lambda-\varpi)\left[1+\left(\frac{R}{r}\right)^2
  \frac{u}{2}\right]+  C\right\}\,\epsilon_\varpi
\nonumber\\
&\quad-  \frac{\mu}{2} \int (u+u')(\Pe'-\Pt')g'\,\epsilon'_\varpi
\,\rmd\Gamma'.
\label{eq:eom-freq}
\end{align}

We note that frequency $\varpi$ in the above equations represents the
temporal variation of the 
neutrino flavor quantum state at given radius $r$, and it shall not be
confused with
the vacuum oscillation frequency $\omega$ which is determined by the
energy of the neutrino.

We also note that the starting point of
collective oscillations is determined by the comparison of the
dispersion in each term of the neutrino propagation
Hamiltonian in Eq.~\eqref{eq:eom-full} with the overall strength
of the neutrino potential. The spreads in the
vacuum Hamiltonian $\Hvac$ and the neutrino potential $\Hnu$ are
dominated by variations in vacuum oscillation frequency $\omega$ and
trajectory parameter $u$, respectively. Therefore, 
we have taken $v_u\approx 1$ for these terms
in Eq.~\eqref{eq:eom-freq} as in Ref.~\cite{EstebanPretel:2008ni}. For
the matter potential 
$\Hmat$, however, the lowest order term of $v_u^{-1} \Hmat$ in the
large-radius expansion is the same
for all neutrinos and does not suppress collective oscillations
\cite{Duan:2005cp}. In Eq.~\eqref{eq:eom-freq} we have included its
next-order term which can suppress collective
oscillations if a very large matter potential is present
\cite{EstebanPretel:2008ni}.

Eq.~\eqref{eq:eom-freq} is the same as that for neutrino
oscillations in the stationary Bulb model except
with replacement $\lambda\rightarrow\lambda-\varpi$.
The flavor-stability analysis of this model has been carried out in
details in Ref.~\cite{Banerjee:2011fj} which we shall not repeat here.
The essence of this analysis is to
find out all the collective oscillation solutions to 
Eq.~\eqref{eq:eom-freq} which are of the form
\begin{align}
\epsilon_\varpi = Q_{\varpi}\, e^{-\rmi\Omega_\varpi r},
\end{align}
where $Q_{\varpi}(\omega,u)$ is independent of $r$,
and $\Omega_\varpi(\lambda,\mu)$ is the collective oscillation frequency.
If
\begin{align}
\kappa_\varpi = \text{Im}(\Omega_\varpi)
\end{align}
is positive, there exists a flavor instability, and
$\epsilon_\varpi$ will grow exponentially in terms of $r$ which can
lead to significant flavor transformation. If there exist multiple
unstable modes, the unstable mode with the largest exponential growth
rate $\kmax$ will eventually dominate.

\section{Results and Discussion}

\begin{figure*}[tb] 
\centering
$\begin{array}{@{}cc@{}}
\includegraphics*[scale=0.58]{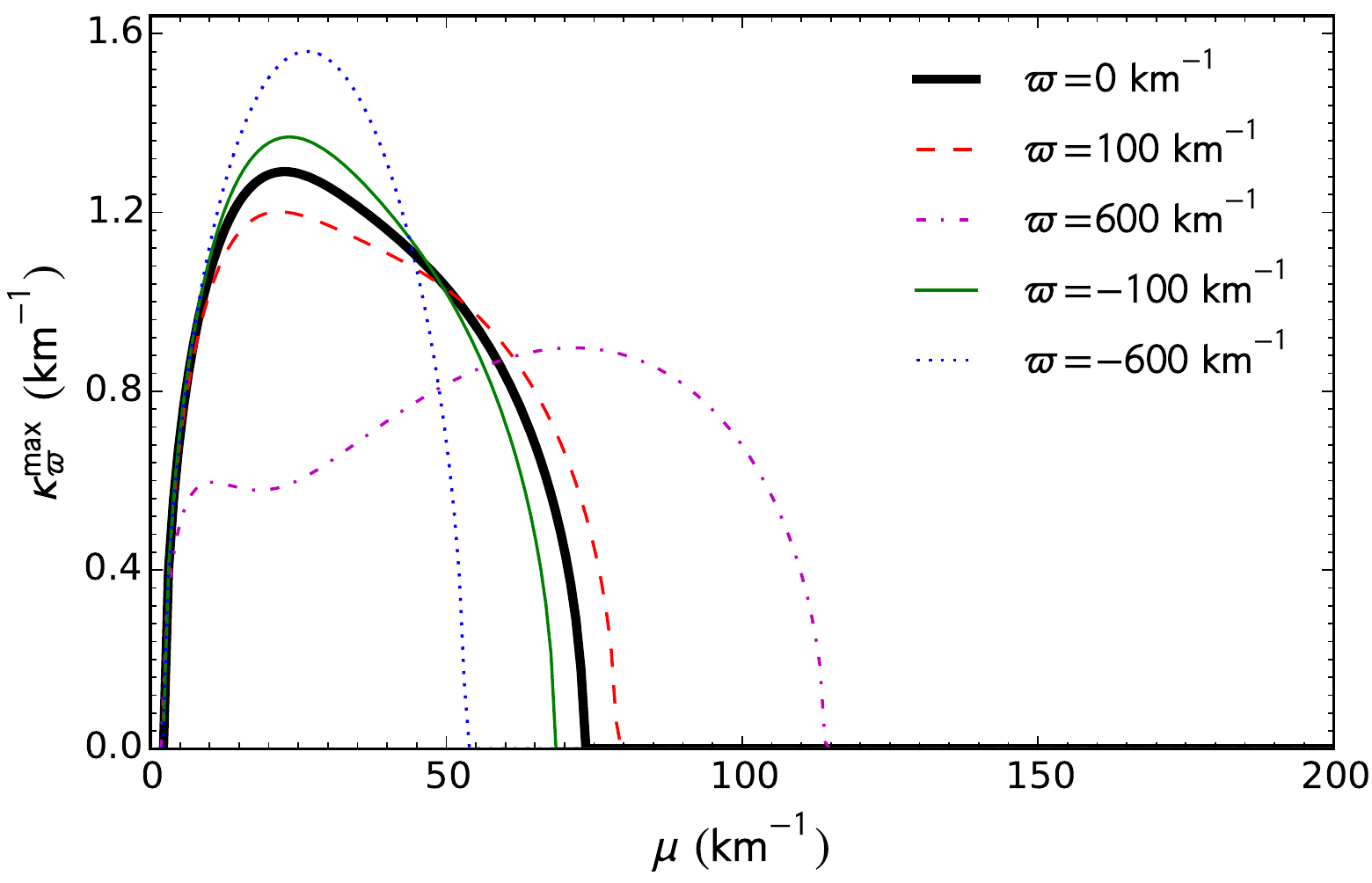}&
\includegraphics*[scale=0.58]{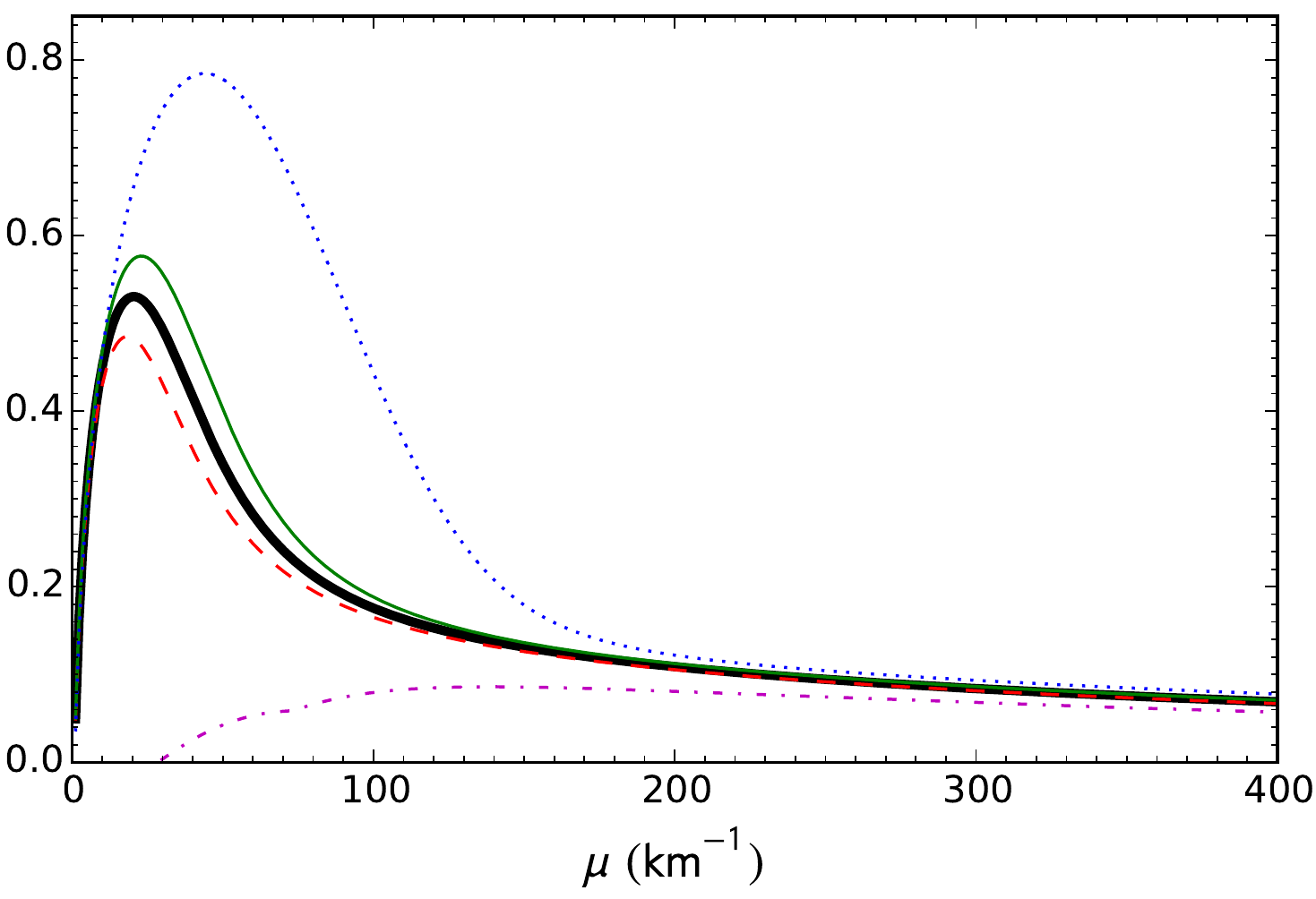}
\end{array}$
\caption{The maximum exponential growth rates $\kmax$ of a few
  Fourier modes with various frequencies $\varpi$ (as labeled) as functions of 
  the strength of the neutrino potential $\mu$ in the neutrino Bulb
  model. The left panel uses a single-energy spectrum 
  \cite{Mirizzi:2015hwa}, and the right panel a continuous spectrum
  \cite{Duan:2010bf,Banerjee:2011fj}. 
  The peaks of $\kappa^\text{max}_{\varpi=0}(\mu)$ correspond to distances
  $107$ km (left) and $149$ km (right) from the center of the SN.
  The matter density is
  assumed to be not large enough to suppress collective neutrino oscillations.}
\label{fig:kappa}
\end{figure*}

We analyzed the flavor instabilities in the time-dependent Bulb model
with two sets of neutrino spectra. In the first case we assume the
same single-energy spectrum as in Ref.~\cite{Mirizzi:2015hwa} in which
all neutrinos and antineutrinos have the same
vacuum oscillation frequency
$\omega_0 =0.68\,\km^{-1}$ and the number fluxes 
are $N_{\nu_e}=1.25\times10^{56}\,\s^{-1}$,
$N_{\bar\nu_e}=8.32\times10^{55}\,\s^{-1}$ and
$N_{\nu_x/\bar\nu_x}=5.20\times10^{55}\,\s^{-1}$.
In the second case we assume the same Fermi-Dirac spectra as 
in Refs.~\cite{Duan:2010bf,Banerjee:2011fj} which have
degeneracy parameters $\eta_{\nu_e}=3.9$, $\eta_{\bar\nu_e}=2.3$ and
$\eta_{\nu_x/\bar\nu_x}=2.1$, average energies
$\langle E_{\nu_{e}} \rangle=9.4\,\MeV$, $\langle E_{\bar\nu_{e}}
\rangle=13.0\,\MeV$, $\langle 
E_{\nu_{x}/\bar\nu_x} \rangle=15.8\,\MeV$, and luminosities
$L_{\nu_{e}}=4.1\times
10^{51}\,\erg\,\s^{-1}$, $L_{\bar\nu_{e}}=4.3\times 10^{51}\,\erg\,\s^{-1}$,
$L_{\nu_{x}/\bar\nu_x}=7.9\times 10^{51}\,\erg\,\s^{-1}$.
In both cases we assume a neutrino sphere of radius $R=10\, \km$ and
mass-squared difference $\Delta
m^2=-2.4\times10^{-3}\,\mathrm{eV}^2$, i.e.\ with an inverted neutrino
mass hierarchy. 

In Fig.~\ref{fig:kappa} we plot
$\kmax$ for a few frequency modes as functions of neutrino potential
strength $\mu$ [see Eq.~\eqref{eq:mu}] assuming that the matter
density is not large enough to suppress collective oscillations
(i.e.\ $v^{-1}\lambda\approx \lambda$ is valid).
In both cases both the instability region and $\kmax$
are about the same for the frequency modes
with $|\varpi|\lesssim 100 \text{ km}^{-1}$.
This is not a coincidence. Compared to the stationary model, the
time-dependent model has a new term in Eq.~\eqref{eq:eom-freq}
\begin{align}
\frac{\varpi}{v_u} \approx
  \varpi +\pfrac{R}{r}^2\frac{u\varpi}{2}.
\end{align}
The first term in the above equation changes only  the real part of
the collective oscillation 
frequency $\Omega_\varpi$ and has no impact on the flavor stability.
The second term depends on the neutrino trajectory and
has a spread $\Delta\varpi\sim (R/r)^2 |\varpi|$. It becomes important only when
\begin{align}
  \Delta\varpi\gtrsim \omega_0,
  \label{eq:comp}
\end{align}
where $\omega_0\sim 1\,\km^{-1}$ is the typical vacuum oscillation
frequency (and also the spread of $\omega$) of supernova neutrinos
with the atmospheric mass-squared 
difference. In both cases collective neutrino oscillations occur at
$r\sim 10 R$ which implies that the stability condition of the frequency
modes with $|\varpi|\lesssim 100 \text{ km}^{-1}$ are about the same.

The above arguments can be generalized to the scenarios where
collective oscillations occur close to the neutrino sphere 
(but not too close so that $r-R\ll R$)
because of, e.g., spatial inhomogeneous oscillation modes
\cite{Duan:2014gfa,Mirizzi:2015fva,Chakraborty:2015tfa,Abbar:2015mca}
or different angular distributions for the neutrino fluxes in different
flavors \cite{Mirizzi:2013rla,Sawyer:2015dsa}.
In these scenarios the spread in $v_u^{-1}\varpi$ is of the same order as
$\varpi$ itself, and
the frequency modes with 
$|\varpi|\lesssim 1 \text{ km}^{-1}\approx (3\,\mus)^{-1}$ should have the
same stability condition.
These arguments also apply in the presence of a large matter
density because the comparison between $\Delta\varpi$ and
$\omega_0$ is not affected by the presence of
the matter potential.

We note that there exists a causality constraint in the time-dependent
Bulb model. Suppose that there is a temporary change in the neutrino fluxes on
one side of the neutrino sphere which lasts for a time interval
$\Delta t$. Because it takes at least $\Delta t'\sim R$ for
this change to propagate throughout the proto-neutron star, the assumption
of the spherical symmetry implies that the inequality $\Delta t\gtrsim
R$ must hold. Therefore, only the oscillation modes of frequencies
\begin{align}
\varpi \lesssim  R^{-1}\sim (10\, \km)^{-1}\approx
(30\,\mus)^{-1}
\end{align}
are allowed in the spherical Bulb model.
From the above discussion we conclude that there should be
no significant difference between the flavor stability conditions in the
time-dependent and stationary Bulb models.
For more general time-dependent SN
models,
collective neutrino oscillations should occur at approximately the
same radius as in the corresponding stationary models unless there
exist very rapid variations in local physical conditions on the timescales
of a few microseconds or shorter.

Meanwhile, the fact that the frequency modes with $|\varpi|\lesssim
1-100\,\omega_0$ all
have similar instability regions also implies that 
the time-translation symmetry can indeed be broken spontaneously by collective
neutrino oscillations in the Bulb model, and that neutrino
oscillations can have a strong time dependence once collective oscillations
begin. As a result, there may exist qualitative differences between
neutrino oscillations in time-dependent and stationary supernova models.

\section*{Acknowledgments}
We thank S.\ Shashank for useful discussions. We
appreciate the 
hospitality of INT/UW where part of this work was done.
This work was supported by DOE EPSCoR grant \#DE-SC0008142 at UNM.

\section*{References}
\bibliographystyle{elsarticle-num}
\bibliography{bulb-time}

\end{document}